\documentclass[aps,prl,twocolumn,floatfix,showpacs]{revtex4}

\usepackage{graphics}
\usepackage{bm}

\def\PRA{{Phys.~Rev.~A} }

\def\JPB{{J.~Phys.~B} }
\def\PRL{{Phys.~Rev.~Lett.} }
\def\RMP{{Rev.~Mod.~Phys.} }
\def\JCP{{J.~Chem.~Phys.} }

\newcommand{\myscalebox}[1]{\scalebox{0.35}[0.35]{#1}}
\newcommand{\myscaleboxa}[1]{\scalebox{0.25}[0.25]{#1}}
\newcommand{\myscaleboxb}[1]{\scalebox{0.22}[0.22]{#1}}
\newcommand{\myscaleboxc}[1]{\scalebox{0.35}[0.22]{#1}}

\newcommand{\be}{\begin{equation}}
\newcommand{\bea}{\begin{eqnarray}}
\newcommand{\eea}{\end{eqnarray}}
\newcommand{\ee}{\end{equation}}

\begin{document}

\title{Probing molecular frame photoionization via laser generated
high-order harmonics from aligned molecules}

\author{Anh-Thu Le,$^1$
R.~R. Lucchese,$^2$, M.~T. Lee,$^3$ and C.~D. Lin$^1$}

\affiliation{$^1$Department of Physics, Cardwell Hall, Kansas
State University, Manhattan, KS 66506, USA\\
$^2$Department of Chemistry, Texas A\&M University, College Station,
Texas 77843-3255, USA\\
$^3$ Departamento de Qu\'imica, Universidade Federal de S\~ao
Carlos, 13565-905, S\~ao Paulo, Brazil}

\date{\today}

\begin{abstract}
Present photoionization experiments cannot measure molecular frame
photoelectron angular distributions (MFPAD) from the outermost
valence electrons of molecules. We show that details of the MFPAD
can be retrieved with high-order harmonics generated by infrared
lasers from aligned molecules. Using accurately calculated
photoionization transition dipole moments for fixed-in-space
molecules, we show that the dependence of the magnitude and phase of
the high-order harmonics on the alignment angle of the molecules
observed in recent experiments can be quantitatively reproduced.
This result provides the needed theoretical basis for ultrafast
dynamic chemical imaging using infrared laser pulses.
\end{abstract}

 \maketitle

Photoionization (PI) is the basic process that allows the most
direct investigation of molecular structure. Measurements of total,
partial and differential PI cross sections have a long history.
Until recently, however, almost all the experiments are performed
from an ensemble of randomly oriented molecules. Thus the rich
dynamical structure of photoelectron angular distribution for
fixed-in-space molecules predicted in the seminal paper by Dill
\cite{Dill76} more than 30 years ago still remains largely
unexplored. In recent years, fixed-in-space PI has been investigated
with x-ray or VUV photons by using the photoelectron-photoion
coincidence technique \cite{weber,liu08}. Following inner-shell or
inner-valence-shell ionization the molecular ion dissociates. The
molecular axis at the time of ionization is inferred from the
direction of motion of the fragment ion if the dissociation time is
short compared to the rotational period. Clearly this method is not
applicable to PI from the highest occupied molecular orbital (HOMO).


Consider photoionization of a linear molecule. For a polarized
light, the differential cross section can be expressed in the
general form \cite{lucchese82} (atomic units are used throughout,
unless otherwise indicated)
\begin{equation}
\frac{d^2\sigma}{d\Omega_{\bm k}d\Omega_{\bm n}}=
\frac{2(2\pi)^3E}{3c}\left|\sum_{lm\mu}I_{lm\mu}Y^*_{lm}(\Omega_{\bm
k})Y^*_{1\mu}(\Omega_{\bm n})\right|^2.
\end{equation}
In this expression, the molecular axis is fixed, the directions of
light polarization and electron emission are given by $\Omega_{\bm
n}$ and $\Omega_{\bm k}$, respectively. This is called the molecular
frame photoionization angular distribution (MFPAD). We calculate the
continuum wavefunction $\Psi^-_{f,klm}$ and the dipole
$I_{lm\mu}=k^{1/2}\langle\Psi_i|r_{\mu}|\Psi^-_{f,klm}\rangle$ (with
$r_{\mu}=z$ for linear polarization) using the iterative Schwinger
variational method \cite{lucchese82} within the
complete-active-space configuration interaction scheme
\cite{lucchese95}. To compare with measurements, integration of
Eq.~(1) over unobserved variables has to be performed, thus losing
much valuable information on the structure of the molecule.

In this Letter we show that recent advances in the generation of
high-order harmonics from molecules by intense infrared lasers has
made it possible to probe fixed-in-space molecular PI from HOMO. Gas
phase molecules can be impulsively aligned by a short sub-picosecond
infrared laser \cite{seideman}. After the pulse is turned off,
molecules will be partially aligned or anti-aligned at the time
intervals of rotational revival \cite{itatani05,kanai05}. During
these revivals which last for tens or hundreds of femtoseconds,
another probe pulse can be used to illuminate molecules to observe
the emission of high-order harmonics. The first experiment of this
type was reported by Kanai {\it et al.} \cite{kanai05} using N$_2$,
O$_2$ and CO$_2$ molecules. Since then many more experiments have
been reported \cite{itatani05,vozzi05,jila07,jila08,riken08,boutu}.
To study the dependence of high-order harmonics generation (HHG) on
the alignment of molecules, the probe laser can be applied at the
different time as the molecules evolve, or by changing the direction
of the probe laser polarization at a fixed time delay. More
recently, HHG from mixed gases \cite{jila07,riken08} and
interferometry technique \cite{jila08} have also been used such that
the phase of the HHG can also be measured.

HHG is a highly nonlinear process and can be understood using the
three-step model \cite{corkum}. Electrons are first released from
the molecule by tunnel ionization and thrown into the laser field.
As the laser's electric field changes direction, the electrons may
be driven back to the ion core after gaining kinetic energy from the
field. When the electrons recombine with the ion, high-order
harmonic photons are emitted. Since the photo-recombination (PR) in
the last step is the time-reversed process of PI, it is thus
expected that HHG spectra hold valuable information on the molecule.
Indeed, HHG spectra had been used to extract the HOMO in N$_2$ using
tomographic method \cite{itatani04}. However the tomographic method
relies exclusively on approximating the continuum electrons by plane
waves which is known to fail in general from the PI studies. In
\cite{atle08,toru08} we have developed a quantitative rescattering
theory (QRS) for HHG where complex HHG dipole moment
$D(\omega,\theta)$ from a fixed-in-space molecule can be expressed
as a product of the transition dipole
$d(\omega,\theta)=\sum_{lm\mu}I_{lm\mu}Y^*_{lm}(\Omega_{\bm
k})Y^*_{1\mu}(\Omega_{\bm n})$ (with ${\bm k}\parallel{\bm n}$) and
the returning electron wave-packet $W(E_k,\theta)$,
\begin{eqnarray}
D(\omega,\theta)=W(E_k,\theta)\times d(\omega,\theta).
\end{eqnarray}
This model establishes the relation between the induced dipole
moment with the transition dipole for PR (or PI). Here $\theta$ is
the alignment angle of the molecular axis with respect to the
polarization direction of the probe laser, and $\omega$=$I_p+E_k$,
with $\omega$ the photon energy of the harmonics, $I_p$ the
ionization potential, and $E_k=k^2/2$ the ``incident" energy of the
returning electron. The validity of this equation has been
established for atomic targets (no $\theta$ dependence for atoms)
\cite{atle08,toru08} as well as for H$_2^+$ target \cite{H2+}. In
both cases,  HHG spectra can be calculated accurately by solving the
time-dependent Schr\"odinger equation (TDSE). In this Letter, we
obtained transition dipole $d(\omega,\theta)$ from the
state-of-the-art PI calculations. The wave-packet $W(E_k,\theta)$ is
obtained from the strong-field approximation
(SFA)\cite{atle08,H2+,lewenstein}. We then use Eq.~(2) to obtain
$D(\omega,\theta)$ for each fixed-in-space molecule. To compare with
experiments, proper coherent convolution with the partial alignment
of molecules has to be carried out [see Eq.~(2) of
Ref.~\cite{jila07}].

\begin{figure}
\mbox{\rotatebox{0}{\myscaleboxa{
\includegraphics{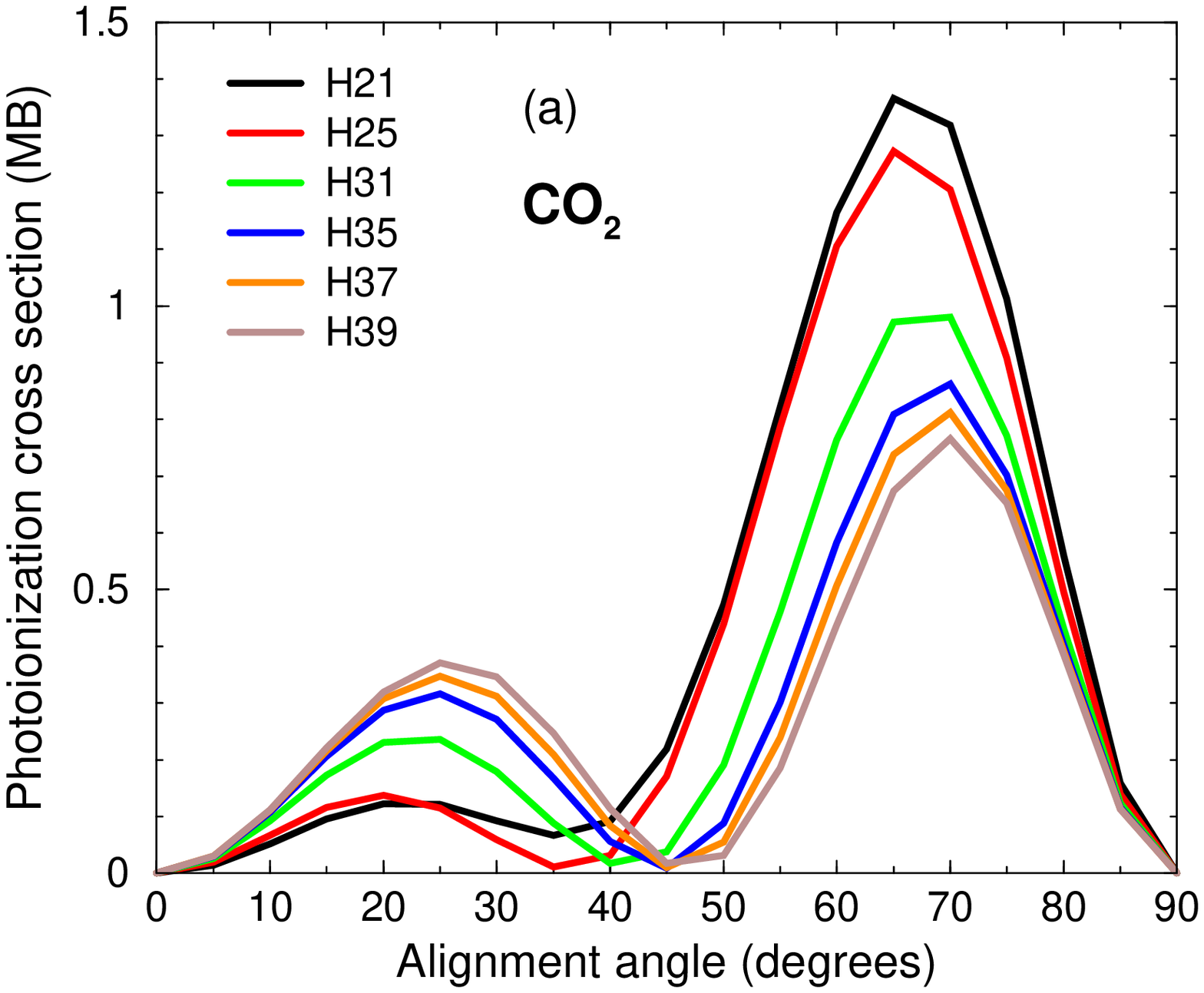}}}}
\mbox{\rotatebox{0}{\myscaleboxa{
\includegraphics{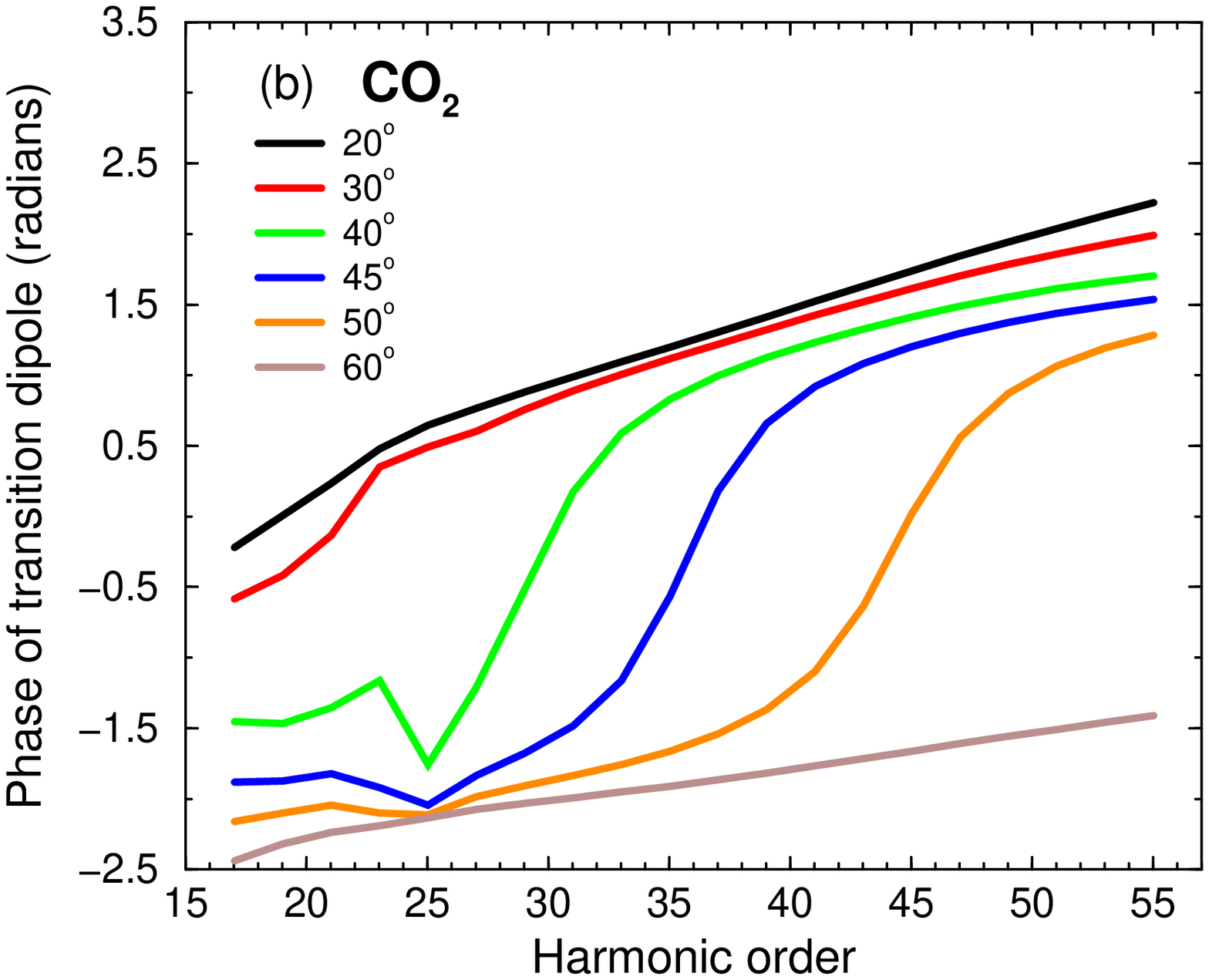}}}}
\mbox{\rotatebox{0}{\myscaleboxa{
\includegraphics{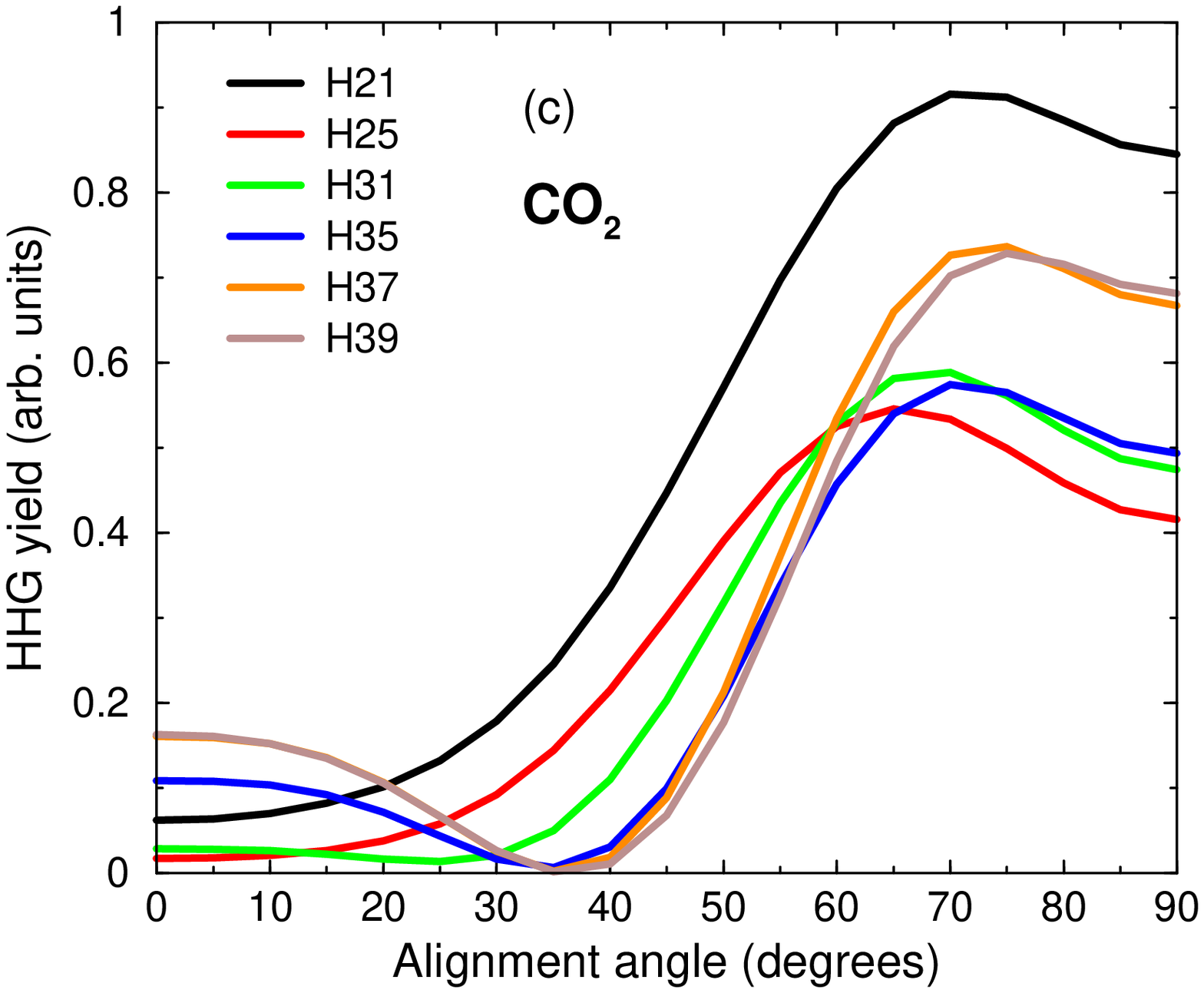}}}}
\mbox{\rotatebox{0}{\myscaleboxa{
\includegraphics{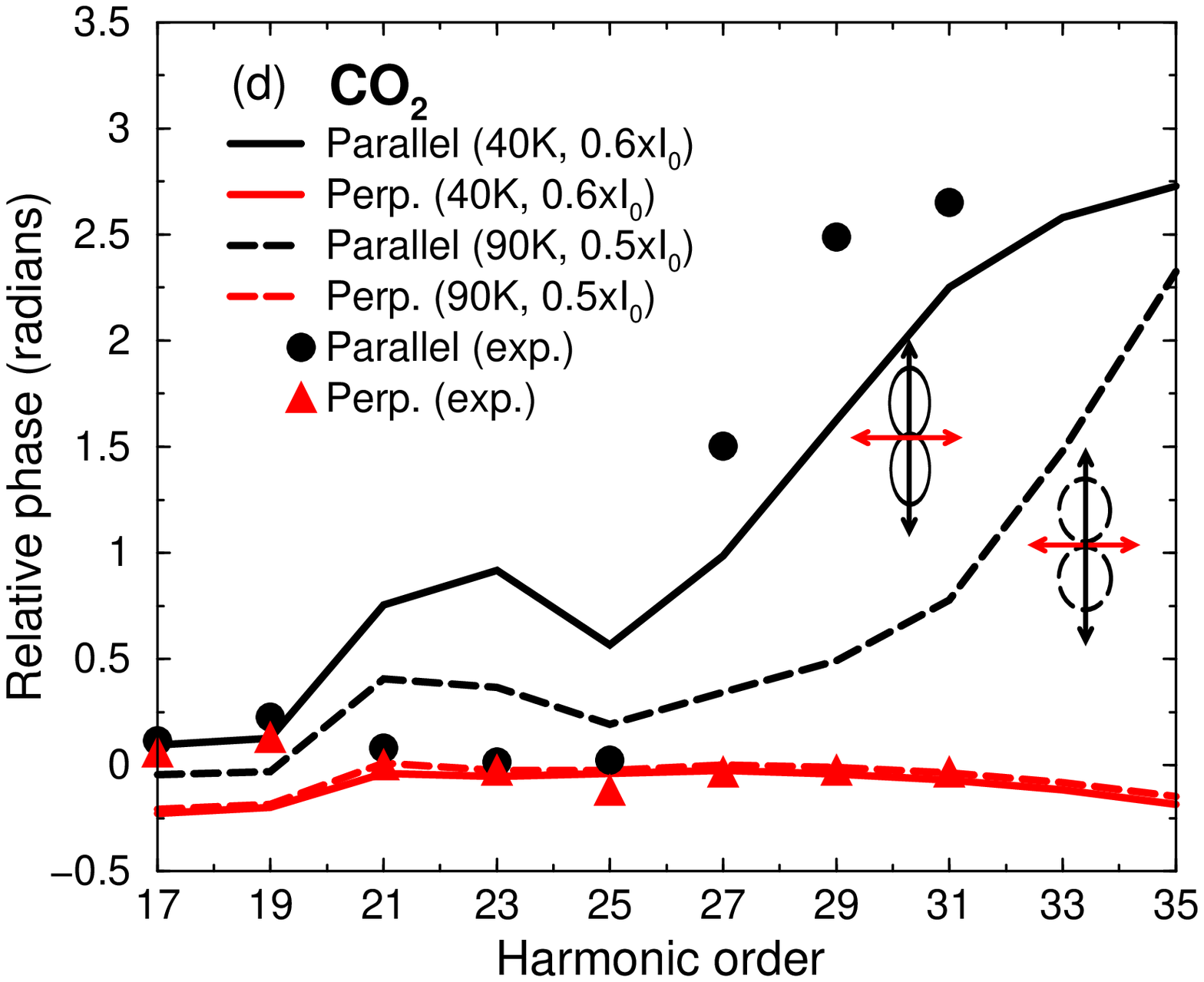}}}}
\caption{(Color online) CO$_2$ differential photoionization cross
section (a) and phase of the transition dipole (b). The photon
energies are expressed in units of harmonic orders for 800-nm laser.
(c) HHG yield as function of angle between pump and probe laser
polarization directions. The laser intensity and duration are of
$0.55\times I_0$ ($I_0=10^{14}$ W/cm$^2$), and $120$ fs for the pump
pulse, and $2.5\times I_0$ and $25$ fs for the probe pulse.
Rotational temperature is taken to be $105$ K. (d) Harmonic phase
(relative to the phase from Kr) for parallel aligned and
perpendicular aligned ensembles, under two sets of parameters (solid
and dashed lines) that lead to two different degrees of alignment
distributions. Experimental data from Boutu {\it el al.}
\cite{boutu} are shown as symbols.} \label{fig1}
\end{figure}

In Fig. 1(a),   we show the theoretical CO$_2$ PI differential cross
section for the emission direction along the light polarization
(${\bm k}\parallel{\bm n}$), which are to be compared to HHG data
below. By changing the orientation of the fixed-in-space molecule,
complementary information on the angular distribution of
photoelectrons from fixed-in-space molecules is obtained. In order
to compare with HHG data, the photon energies are expressed in units
of photon energy of 800-nm laser ($1.55$ eV). First, we note that
the differential cross section is large when the molecule is aligned
at a large angle. The cross sections vanish at $\theta=0$ and
$\pi/2$ due to the $\pi_g$ symmetry of the HOMO and the dipole
selection rule for the final state. Second, the cross section shows
near-zero minima for angles around 35$^{\circ}$ to 45$^{\circ}$ for
harmonics above the 25th order, or H25. The position of the minimum
moves to a larger angle, as energy increases. These minima have been
interpreted as evidence of interference of electron waves from the
atomic centers. In Fig.~1(b), we show the phase of the PI dipole vs
photon energy, for alignment angles from 20$^{\circ}$ to
60$^{\circ}$. Between 30$^{\circ}$ and 45$^{\circ}$ the dipole phase
undergoes rapid change between H25 and H39. This is the region where
the cross sections are near the minima [see Fig.~1(a)]. The phase
change is about 2.0 to 2.5, instead of $\pi$ as should be expected
if the cross section indeed goes to zero. For angles below
20$^{\circ}$ and above 60$^{\circ}$ the phase evolves smoothly vs
photon energy. The ``rich" structures in the theoretical amplitude
and phase of $d(\omega,\theta)$  shown in Fig.~1 have never been
observed directly in PI experiments. Below we show that they have
been observed in recent HHG measurements.

In Fig.~1(c) we show the simulated typical HHG yields from aligned
CO$_2$, as a function of the angle between the pump and probe lasers
polarization directions. These results are consistent with recent
experiments \cite{boutu,mairesse08,JILA-private}, and with earlier
theoretical results \cite{nalda04,atle06,atle07}. The results also
resemble the data for the induced dipole retrieved from mixed gases
experiments by Wagner {\it et al.} \cite{jila07} (see their Fig.~4).
In our simulation, the laser parameters are taken from the
experiment of \cite{jila08}.  The alignment distribution is obtained
from numerical solution of the TDSE  within the rotor model
\cite{seideman,atle06,atle07}. At the delay time corresponding to
the maximum alignment near half-revival (1/2 of the rotational
period), a probe beam with polarization direction varying from 0 to
$90^{\circ}$ is used to generate high-order harmonics. Comparing
Fig.~1(c) with Fig.~1(a), we note that the HHG yields follow the
general angular and photon energy dependence of the the differential
PI cross sections. Both figures show large yields at large angles,
and minima near 35$^{\circ}$ for harmonic orders above H31. Due to
the averaging over the molecular alignment distributions, the
angular dependence of HHG is smoother.

We next show that the phase of fixed-in-space PI transition dipole
can also be probed by comparing to the phase of the harmonics.
Experimentally the harmonics phase can be extracted from
measurements of HHG using mixed gases \cite{boutu,riken08} or
interferometry method \cite{jila08}. In Fig.~1(d),  we show the
recent experimental data of Boutu {\it et al.} \cite{boutu} where
the harmonics phases (relative to that from Kr) are obtained for the
parallel aligned and perpendicularly aligned ensembles, shown by
black and red symbols, respectively. For the latter, the phase does
not change much within H17 to H31. For the parallel aligned
molecules, the phase jump from H17 to H31 was reported to be $2.0\pm
0.6$ radians (in the figure it was shown at 2.6 radians). Our
simulation results are shown for two different ensembles, with
alignment distributions confined in a cone angle of $~25^{\circ}$
and $35^{\circ}$ at half maximum, resulting from different pump
beams and gas temperatures (see Caption). From Fig.~1(b), we note
that for a fixed harmonic, the phase from the parallel aligned
ensemble is larger if the angular spread of the molecular
distributions is smaller. Therefore, the position of the phase jump
for the less aligned ensemble (dashed line) slightly shifted toward
higher energies. For the perpendicularly aligned molecules, the
phase is small and does not change much with the harmonic order.
This is consistent with the phase shown in Fig.~1(b). Thus the phase
of HHG in Fig. 1(d) is seen to mimic the phase of the PI dipole
shown in Fig. 1(b).

\begin{figure} \mbox{\rotatebox{0}{\myscalebox{
\includegraphics{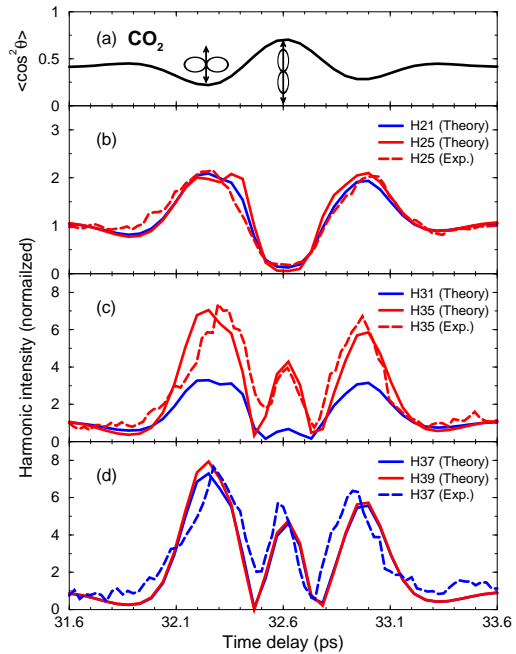}}}}
\caption{(Color online) Normalized  HHG yield (vs isotropically
distributed molecules) from CO$_2$ for different harmonics as
function of pump-probe delay time near 3/4-revival. Laser parameters
are the same as in Fig.~1(c). The experimental data are taken from
Zhou {\it et al} \cite{jila08}. The alignment parameter
$\langle\cos^2\theta\rangle$ is plotted for reference (a).}
\label{fig2}
\end{figure}

Next we show in Fig.~2 how the HHG yields change with the time delay
near the  3/4-revival where the molecule can be most strongly
aligned. Our simulations were carried out with the laser parameters
from Zhou {\it et al.} \cite{jila08}. The degree of alignment (top
panel), as measured by $\langle\cos^2\theta\rangle$ is maximum when
molecules are maximally aligned (vertically in the figure) and
minimum when molecules are anti-aligned. Our results are shown for
the same harmonics that had been analyzed in \cite{jila08} (see
their Fig.~2 - according to a new assignment by the authors of
\cite{jila08}, the harmonic orders should be properly shifted down
by two orders, compared to the ones given in their original paper
\cite{JILA-private}). For the lower harmonics H21 and H25, the
yields follow the inverse of the alignment parameter. This is easily
understood from Fig.~1(a) which shows that the PI cross sections at
large angles are much larger than at small angles. For the higher
orders, Fig.~1(a) indicates that the PI cross sections show two
humps, with the one at smaller angles only a factor of about two to
three times smaller than the other. Qualitatively, this explains why
the HHG yields for H31 and up show a pronounced peak for the
parallel alignment. We note a quantitatively good agreement between
our calculations and the experimental data.

Evolution of the HHG yields has also been studied by Wagner {\it et
al.} \cite{jila07} using the mixed gases technique. The yield and
phase for H31 vs time delay near half-revival are shown in Fig.~3(c)
and (b), respectively. Our simulations (solid lines) are also in
good agreement with these measurements. The HHG yield is small when
the molecules are maximally aligned. This is in agreement with H31
seen in Fig.~2(c) for the delay time near 3/4-revival. However, we
note that the phase is maximum when the molecules are maximally
aligned. This can be understood from the phase in Fig.~1(b) where it
shows that the phase is large when the alignment angle is small. It
also shows that the phase is small when the alignment angle is
large, thus for anti-aligned molecules the phase of the HHG should
be small, as seen in the experimental data and in the simulation.

We remark that the laser intensity used in the JILA experiments is
quite high, thus the effect of the depletion of the ground state by
the laser should be considered \cite{atle06,atle07,xu08}. For
CO$_2$, the alignment dependence of the ionization rate is still not
fully settled yet. The results from the MO-ADK theory \cite{moadk},
the SFA theory and the experiment \cite{NRC-exp} all disagree with
each other, with the SFA predicting a peak near $40^{\circ}$,
compared to $30^{\circ}$ from MO-ADK theory, and $45^{\circ}$ from
experiment \cite{NRC-exp}. Using ionization rates from MO-ADK, we
have not been able to simulate experimental data accurately. We thus
used ionization rates from SFA, except that we renormalize the SFA
rate to that of the MO-ADK rate at laser intensity of $I_0$, which
gives a factor of $10$. Note that the same correction factor has
been found for the SFA ionization from Kr, which has almost the same
ionization potential as for CO$_2$. With the corrected SFA rate, we
found that our simulations give a better quantitative agreement with
the JILA data \cite{jila08}.

\begin{figure}
\mbox{\rotatebox{0}{\myscaleboxc{
\includegraphics{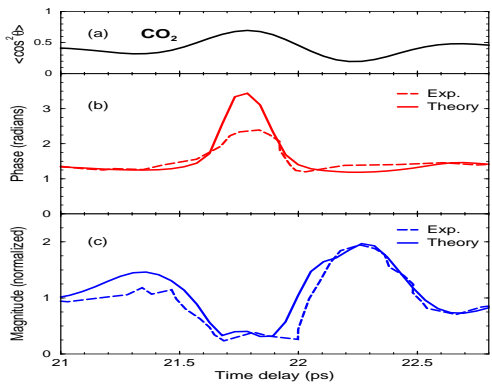}}}}
\caption{(Color online) Alignment parameter
$\langle\cos^2\theta\rangle$ (a), phase (b) and magnitude (c) of the
31th order harmonic from CO$_2$, as functions of pump-probe delay
time near half-revival. The experimental data are taken from Wagner
{\it et al} \cite{jila07}. The laser intensity and duration are of
$0.38\times I_0$ and $140$ fs for the pump, and $2.5\times I_0$ and
$30$ fs for the probe. Rotational temperature is taken to be $70$
K.} \label{fig3}
\end{figure}

We have also studied N$_2$ molecules. The fixed-in-space
differential PI cross sections are shown in Fig.~4(a). For H17 to
H21 the cross sections are quite large since this is in the tail
region of the famous shape resonance in N$_2$. For these orders, the
cross sections are highly forwardly peaked. For energies above H25,
the cross sections become much smaller, and they are of the same
order of magnitude between parallel and perpendicular alignments.
However, note that the cross sections have minima at large angles
around 60$^{\circ}$ -- 70$^{\circ}$. The (normalized) HHG spectra of
N$_2$ has been measured vs the time delay in \cite{itatani05}. Their
results are shown in Fig.~4(c), along with the results from our
simulation. Clearly for H21-H25, the yield (normalized to the
isotropic distribution case) is much larger than that from the
higher harmonics when the molecules are maximally aligned, which is
consistent with the PI cross sections shown in Fig.~4(a). Once
again, the experimental HHG spectra can be reproduced within the QRS
based on the accurate dipole transition amplitudes from PI
calculation, thus allowing us to probe the MFPAD for HOMO.

\begin{figure}
\mbox{\rotatebox{0}{\myscaleboxa{
\includegraphics{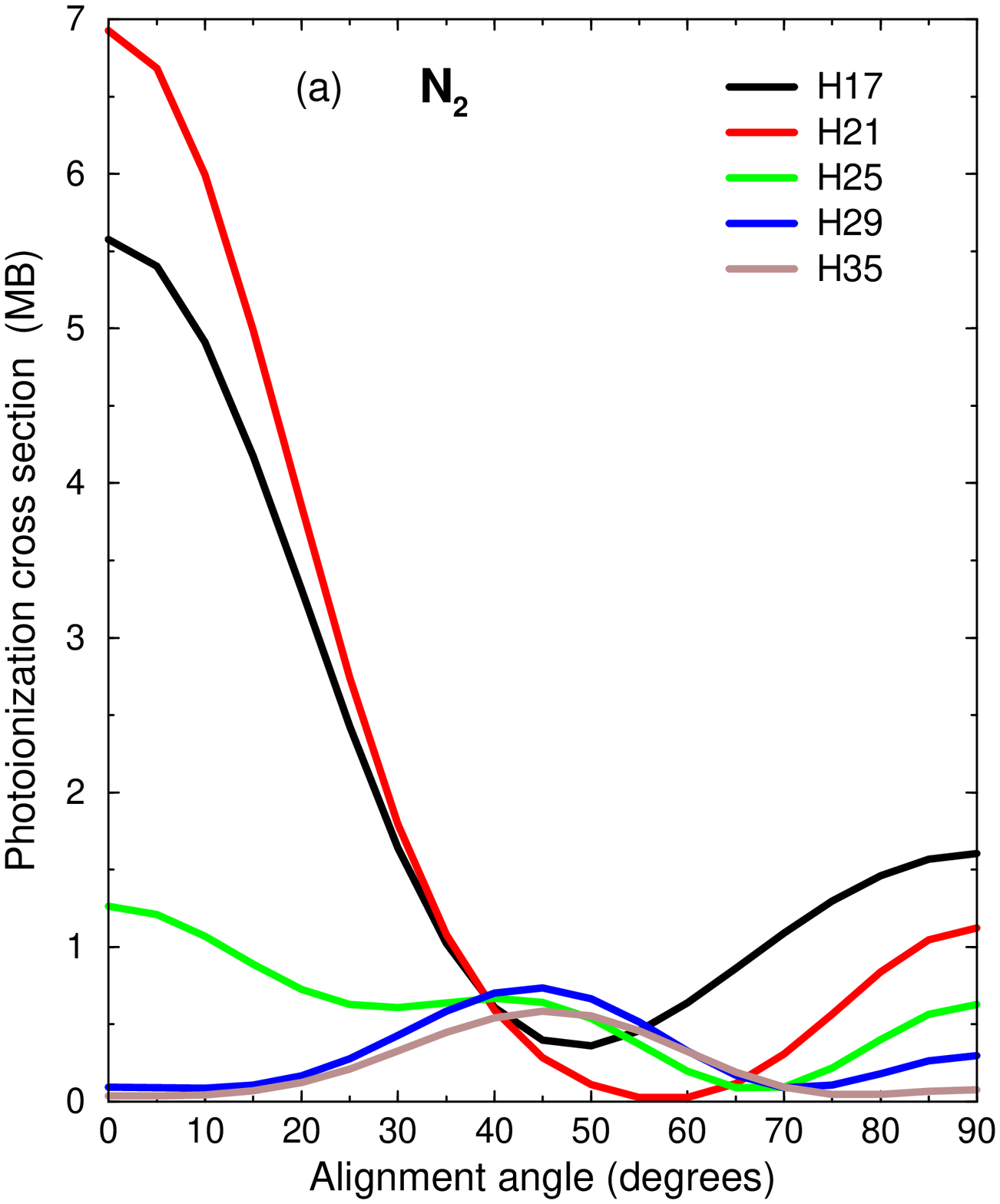}}}}
\mbox{\rotatebox{0}{\myscaleboxb{
\includegraphics{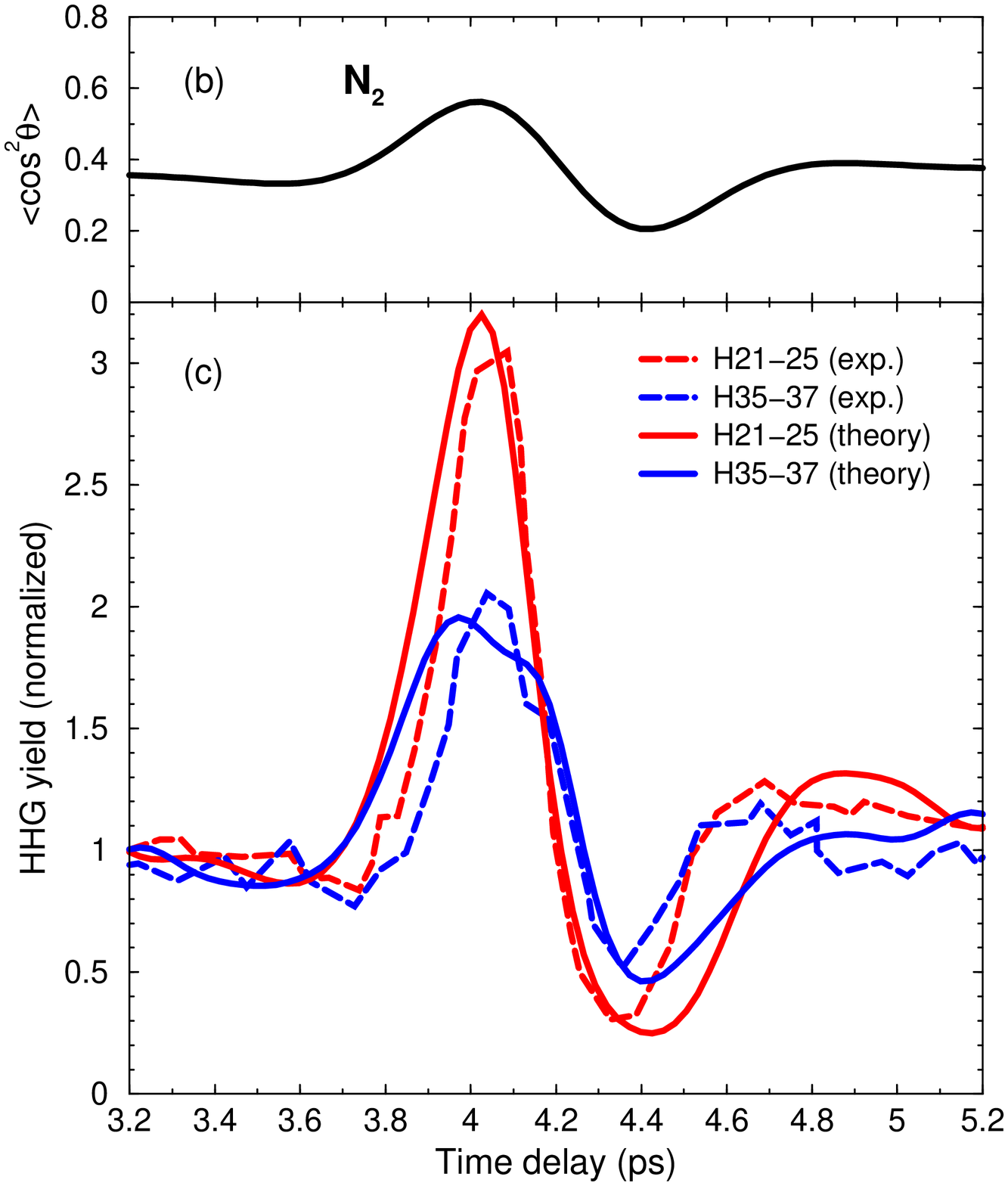}}}}
\caption{(Color online) (a) Same as in Fig.~1(a), but for N$_2$. (b)
and (c): Alignment parameter $\langle\cos^2\theta\rangle$ and
harmonic yields H21-H25 and H35-H37 from N$_2$. The experimetal data
and laser parameters are taken from Itatani {\it et al}
\cite{itatani05}. } \label{fig4}
\end{figure}

In conclusions, we have shown that molecular frame photoionization
can be probed directly using laser-generated high-order harmonics
from aligned molecules. This method is particularly useful for
probing fixed-in-space molecular photoionization from the HOMOs.
Alternatively, using accurate PI dipole matrix elements calculated
from the state-of-the-art molecular photoionization codes, we have
illustrated that the nonlinear HHG spectra can be accurately
calculated [see Eq. (2)]. By varying laser intensity, HHG spectra
from  lower-lying orbitals can also be similarly probed
\cite{stanford}. Having established the connection between HHG and
the PI dipole matrix elements, see Eq.~(2), time-resolved PI cross
sections can be extracted from HHG spectra in a typical pump-probe
experimental setup, thus opening up the opportunity of using
infrared laser pulses for ultrafast dynamic chemical imaging.

We thank X. Zhou, N. Wagner, M. Murnane, H. Kapteyn, and P. Salieres
for communicating their results to us and the valuable discussions.
This work was supported in part by the Chemical Sciences,
Geosciences and Biosciences Division, Office of Basic Energy
Sciences, Office of Science, U. S. Department of Energy.

\end{document}